%% file: TOP2018_ATLAS_CMS_xsec.tex
\documentclass[12pt]{article}
\usepackage{graphicx}
\usepackage{xspace}
\usepackage{amsmath}
\newcommand\pubnumber{SNSN-323-63}
\newcommand\pubdate{\today}

\def\institute{
	Deutsches Elektronen-Synchrotron (DESY) \\
	Notkestr. 85, 22607 Hamburg, GERMANY
	}

\def\Title#1{\begin{center} {\Large #1 } \end{center}}
\def\Author#1{\begin{center}{ \sc #1} \end{center}}
\def\Address#1{\begin{center}{ \it #1} \end{center}}

\newcommand\pubblock{\rightline{\begin{tabular}{l} \pubnumber\\
         \pubdate  \end{tabular}}}
\newenvironment{Abstract}{\begin{quotation}  }{\end{quotation}}
\newenvironment{Presented}{\begin{quotation} \begin{center} 
             PRESENTED AT\end{center}\bigskip 
      \begin{center}\begin{large}}{\end{large}\end{center} \end{quotation}}

\input econfmacros.tex

\begin{document}
\begin{titlepage}

\pubblock

\vfill
\Title{Measurements of the inclusive $\rm t\bar{t}$ production cross section \\ at the ATLAS and CMS experiments}
\vfill
\Author{ Matteo M. Defranchis \\ on behalf of the ATLAS and CMS Collaborations} %\support}
\Address{\institute}
\vfill
\begin{Abstract}

%I summarize the results of the most recent measurements of the inclusive $\mathrm{t\bar{t}}$ cross section performed with the ATLAS and CMS experiments at the CERN LHC. These include results obtained using proton-proton collision data at the centre-of-mass energies of 5.02, 8, and 13~TeV, and the first results in proton-lead collisions, at the nucleon-nucleon centre-of-mass energy of 8.16~TeV. A new result at $\sqrt{s} = 13~\mathrm{TeV}$ by the CMS Collaboration obtained with 35.9~fb$^{-1}$ of proton-proton collisions  was presented at this conference for the first time. The total cross section is determined both for a fixed top quark mass value in the simulation of $m_{\mathrm{t}}^{\mathrm{MC}} = 172.5~\mathrm{GeV}$, and simultaneously with $m_{\mathrm{t}}^{\mathrm{MC}}$.

The results of the most recent measurements of the inclusive $\mathrm{t\bar{t}}$ cross section performed by the ATLAS and CMS experiments at the CERN LHC are summarized. These include results obtained in proton-proton collisions at different centre-of-mass energies, and the first observation of $\mathrm{t\bar{t}}$ production in proton-lead collisions. A new result at $\sqrt{s} = 13~\mathrm{TeV}$ by the CMS Collaboration is presented for the first time at this conference. In this analysis, the total cross section is determined both for a fixed top quark mass value in the simulation of $m_{\mathrm{t}}^{\mathrm{MC}} = 172.5~\mathrm{GeV}$, and simultaneously with $m_{\mathrm{t}}^{\mathrm{MC}}$.

\end{Abstract}
\vfill
\begin{Presented}
$11^\mathrm{th}$ International Workshop on Top Quark Physics\\
Bad Neuenahr, Germany, September 16--21, 2018
\end{Presented}
\vfill
\end{titlepage}
\def\thefootnote{\fnsymbol{footnote}}
\setcounter{footnote}{0}

\section{Introduction}

Measurements of the top quark-antiquark (\ttbar) production cross section (\stt) have been performed with increasing precision at the ATLAS~\cite{Aad:2008zzm} and CMS~\cite{Chatrchyan:2008aa} experiments at several centre-of-mass energies ($\sqrt{s}$) in different decay channels. Theoretical predictions for \stt can be derived in quantum chromodynamics (QCD) at next-to-next-to-leading order (NNLO) in perturbation theory. The precision of the calculation is further improved by including next-to-next-to-leading logarithmic (NNLL) corrections~\cite{Czakon:2011xx}. 
The uncertainty in the predicted \stt ranges from 6 to 7\%, depending on the centre-of-mass energy, while experimental measurements can reach a precision of 3 to 4\%.  \\%Measurements of \stt can therefore probe the validity of perturbative QCD calculations. \\

The \ttbar production cross section depends on fundamental parameters of the standard model (SM), in particular the strong coupling constant (\as) and the top quark mass (\mt), and on the parton distribution functions (PDF) of the proton. Measurements of \stt can therefore be used to constrain SM parameters, but also to probe models of physics beyond the SM. %At the LHC, over 10 \ttbar pairs are produced every second at the current centre-of-mass energy of 13~TeV, which provides a unique opportunity to study this process in detail.
\\
%At the current centre-of-mass energy of 13~TeV, over 10 \ttbar pairs are produced every second at the nominal instantaneous luminosity, making the Large Hadron Collider (LHC) a suitable environment to study this process in detail.\\

Typically, the procedure to measure \stt consists of two separate steps. First, a visible cross section (\sttvis) is measured in the fiducial phase space, where systematic uncertainties can be constrained. %This is estimated as the number of signal events in data divided by the total integrated luminosity and corrected for the efficiency of the event selection. 
The visible cross section is then extrapolated to the full phase space by correcting for the acceptance of the event selection, which is estimated using simulation. %The most precise measurements are obtained in the lepton+jets and  dilepton channels, which have the advantage of a high signal purity.

%\section{ATLAS and CMS measurements at 7 and 8 TeV}

%\section{ATLAS and CMS measurements at 13~TeV}

%\section{CMS measurement at 5.02~TeV and first observation of \ttbar production in p-Pb collisions at 8.16~TeV}

\section{Recent ATLAS and CMS measurements}

%The measurements of \stt performed at the ATLAS and CMS experiments using proton-proton collisions at  centre-of-mass energies of 7, 8, and 13~TeV are summarized in Fig.~\ref{fig:xsec_7_8_13}. The most precise measurements are obtained in the \emu and lepton+jets channels, which have the advantage of a high signal purity. The precision achieved by the two experiments is similar at all centre-of-mass energies. In the following, a selection of the most recently published measurements is presented.\\

\begin{figure}[!h]
	\centering
	\includegraphics[width=.32\textwidth]{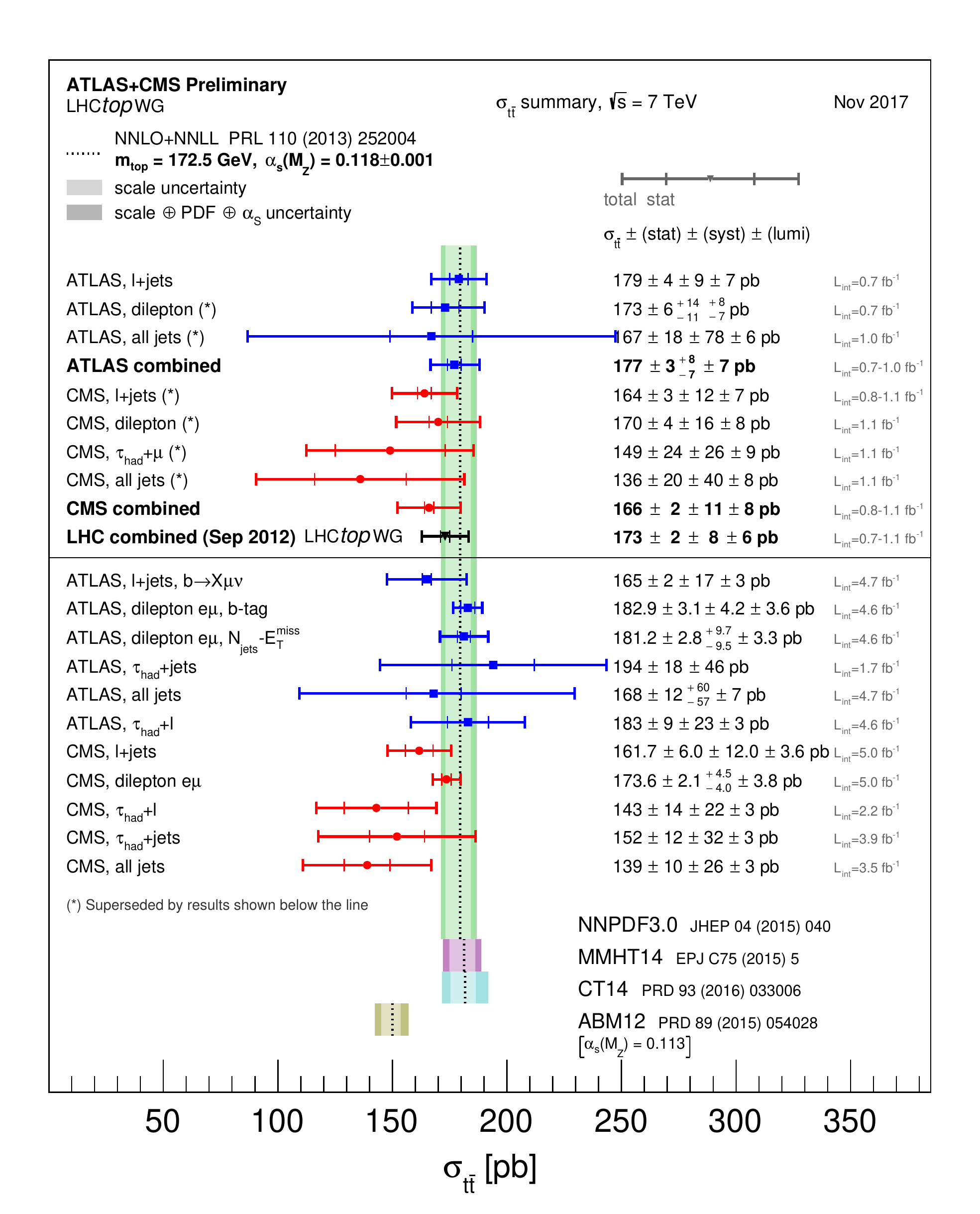}
	\includegraphics[width=.32\textwidth]{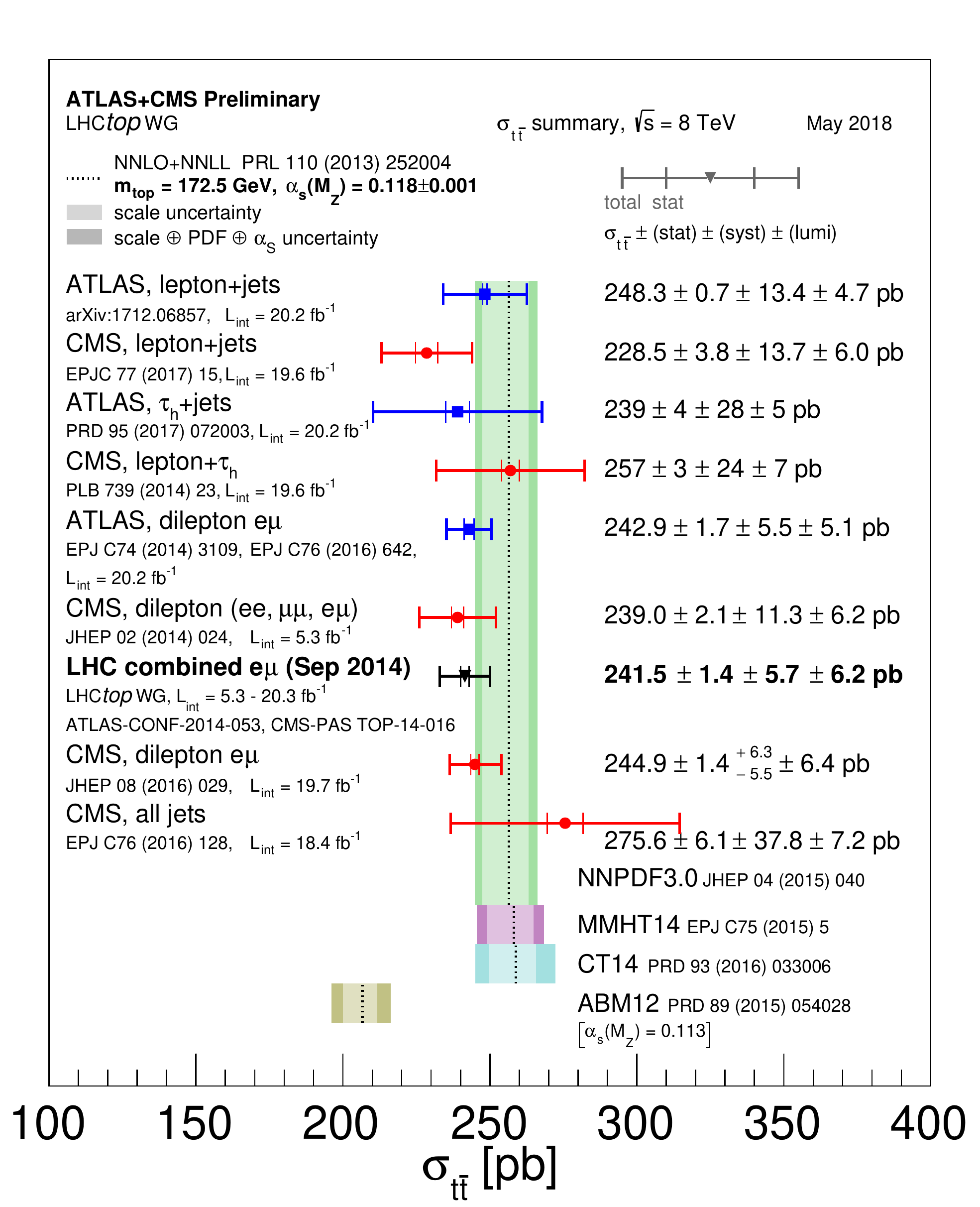}
	\includegraphics[width=.32\textwidth]{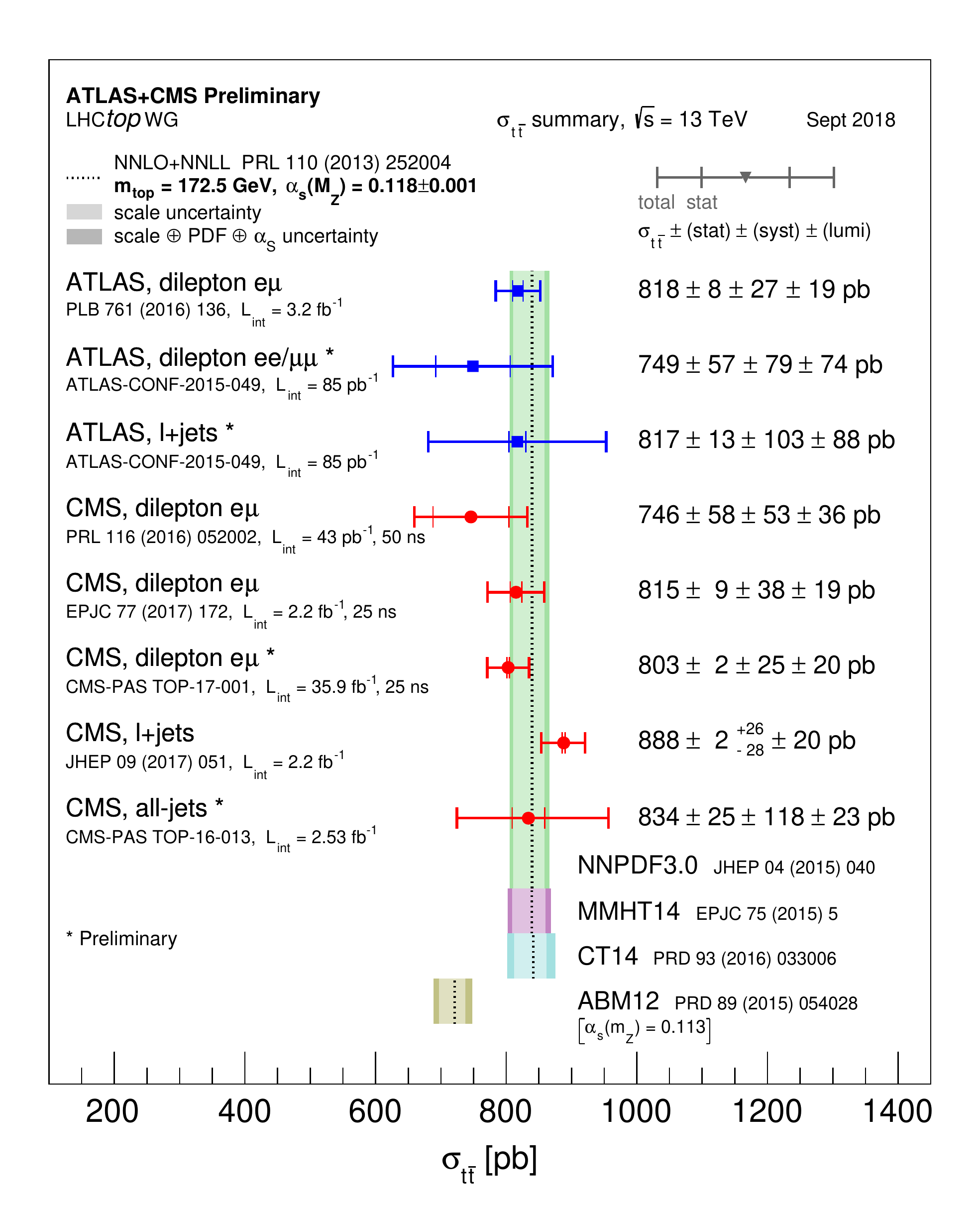}
	\caption{Measurements of the inclusive \ttbar cross section by the ATLAS and CMS Collaborations at 7 (left), 8 (middle), and 13~TeV (right)~\cite{sum-plots}.}
	\label{fig:xsec_7_8_13}
\end{figure}

The measurements of \stt performed at the ATLAS and CMS experiments using proton-proton collisions at  centre-of-mass energies of 7, 8, and 13~TeV are summarized in Fig.~\ref{fig:xsec_7_8_13}. The most precise measurements are obtained using final states having one electron and one muon of opposite charge (\emu) or a charged lepton and additional jets (lepton+jets), which have the advantage of a high signal purity.  \\

%In the following, a selection of the most recently published measurements is presented.\\

A recent result at $\sqrt{s} = 8~\mathrm{TeV}$ by the ATLAS Collaboration is obtained using events in the lepton+jets channel, with an integrated luminosity of 20.2~fb$^{-1}$~\cite{Aaboud:2017cgs}. In this analysis, the \ttbar cross section is determined simultaneously with the b~tagging efficiencies and a global jet energy scale factor. Selected events are split into three different signal regions based on the jet and b-tagged jet multiplicities. In categories with higher background contamination, a neural network is trained to separate the signal from the backgrounds, while in the category with the highest signal purity (that corresponding to four selected jets, two of which are b-tagged), the invariant mass of the two non-b-tagged jets is used to constrain the jet energy scale factor, exploiting the constraint from the mass of the reconstructed W~boson.
In addition, the event categorization is exploited to constrain the b~tagging efficiency. The resulting value of the cross section is $\stt = 248.3 \pm 0.7 \stat \pm 13.4 \syst \pm 4.7 \lum \pb$. The uncertainties denote the statistical uncertainty, the systematic uncertainty, and that coming from the uncertainty in the integrated luminosity. The  relative uncertainty of 5.7\% is dominated by the uncertainty in the gluon PDF at high momentum fraction. \\
%, which affects the extrapolation to the full phase space. \\

A precise measurement at $\sqrt{s} = 13~\mathrm{TeV}$ was performed by the CMS Collaboration in the lepton+jets channel, with an integrated luminosity of 2.2~fb$^{-1}$~\cite{Sirunyan:2017uhy}. The cross section is determined by means of a likelihood fit where the systematic uncertainties are treated as nuisance parameters and constrained using the data. Selected events are classified in 44 orthogonal categories of jet and b-tagged jet multiplicities, lepton charge and lepton flavour. The \mlb distribution, i.e. the minimum invariant mass between the lepton and a reconstructed b~jet, is used to discriminate the signal from the backgrounds. 
%The dependence of the \mlb distribution on the top quark mass in the simulation is taken into account in the likelihood.
The total cross section is measured to be $\stt  =  888 \pm 2 \stat \pm^{26}_{28} \syst \pm 20 \lum \pb$. The relative uncertainty of 3.7\% is dominated by the uncertainty in the b~tagging efficiency and in the normalization of the W+jets background. The resulting cross section is also used to determine the pole mass of the top quark, which is found to be $170.6 \pm 2.7 \GeV$. \\

A precise measurement at $\sqrt{s} = 13~\mathrm{TeV}$ was also performed by the ATLAS Collaboration, using events in the \emu final state corresponding to an integrated luminosity of 3.2~fb$^{-1}$~\cite{Aaboud:2016pbd}. In this analysis, the cross section is determined simultaneously with the b~tagging efficiency. Events with exactly one and two b-tagged jets are selected and split into categories of b-tagged jet multiplicity. %The number of events in each category is expressed in terms of the b~tagging efficiency, the event selection efficiency, the total cross section and the residual correlation in the b~tagging of two jets.
The total cross section is measured to be $\stt = 818 \pm 8 \stat \pm 27 \syst \pm 19 \lum \pm 12~(\mathrm{beam}) \pb$, with a relative uncertainty of 4.2\%. The result is used to perform a precise determination of the ratio of \stt to the Z boson production cross section~\cite{Aaboud:2016zpd}, which benefits from the cancellation of several relevant systematic uncertainties. This quantity is sensitive to the ratio of the gluon to the quarks PDF in the proton, and it was demonstrated that the result can be used to improve the precision in PDF determinations~\cite{Aaboud:2016zpd}. \\

The first measurement of the \ttbar cross section at $\sqrt{s} = 5.02~\mathrm{TeV}$ was recently performed by the CMS Collaboration using data collected in a dedicated LHC run in 2015, corresponding to an integrated luminosity of $27.4~\mathrm{pb^{-1}}$~\cite{Sirunyan:2017ule}. Events in the \emu, \mumu and lepton+jets final states are considered. A cut-and-count method is used for events in the dilepton channels, whereas a fit to categories of b-tagged jet multiplicity is used for the lepton+jets channel. The cross section is measured to be $\stt = 69.5 \pm 6.1 \stat \pm 5.6 \syst \pm 1.6 \lum \pb$, in good agreement with the NNLO theoretical prediction $\stt^{\mathrm{NNLO}} = 68.9 \pm^{1.9}_{2.3}(\mathrm{scale}) \pm 2.3~(\mathrm{PDF}) \pm^{1.4}_{1.0}(\alpha_\mathrm{S}) \pb$. Despite the fact that the measurement is limited by the statistical uncertainty, it was demonstrated that the result can be used to constrain the gluon PDF uncertainty at high momentum fraction~\cite{Sirunyan:2017ule}. \\

An important result by CMS is the observation of \ttbar production in proton-lead collisions~\cite{Sirunyan:2017xku}. The analysis is performed with $174~\mathrm{nb}^{-1}$ of data at a nucleon-nucleon centre-of-mass energy of 8.16~TeV. Lepton+jets events are selected and classified in categories of b-tagged jet multiplicity. The cross section is determined from a likelihood fit to the invariant mass of jets originated from the decay of W~bosons, simultaneously with  the b~tagging efficiency and a global jet energy scale factor. The significance of the \ttbar signal is found to be above five standard deviations.

%\section{Observation of \boldmath \ttbar production in p-Pb collisions}

\section{CMS results at 13~TeV with \boldmath \lumiv}

A new measurement at $\sqrt{s} = 13~\mathrm{TeV}$ is presented for the first time at this conference by the CMS Collaboration~\cite{TOP-17-001}. This is the first result obtained using data collected during the 2016 LHC run, corresponding to an integrated luminosity of 35.9~fb$^{-1}$. The measurement is performed using two different approaches: in the first, the cross section is determined at the fixed top quark mass in the simulation of $\mtmc = 172.5~\GeV$, using events in all the dilepton channels (\emu, \mumu, \ee). In the second approach, the cross section is measured simultaneously with \mtmc, and is therefore determined at the optimal mass point. For simplicity, only the \emu final state is considered in this case. \\

\begin{figure}[!h]
	\includegraphics[width=0.5\textwidth]{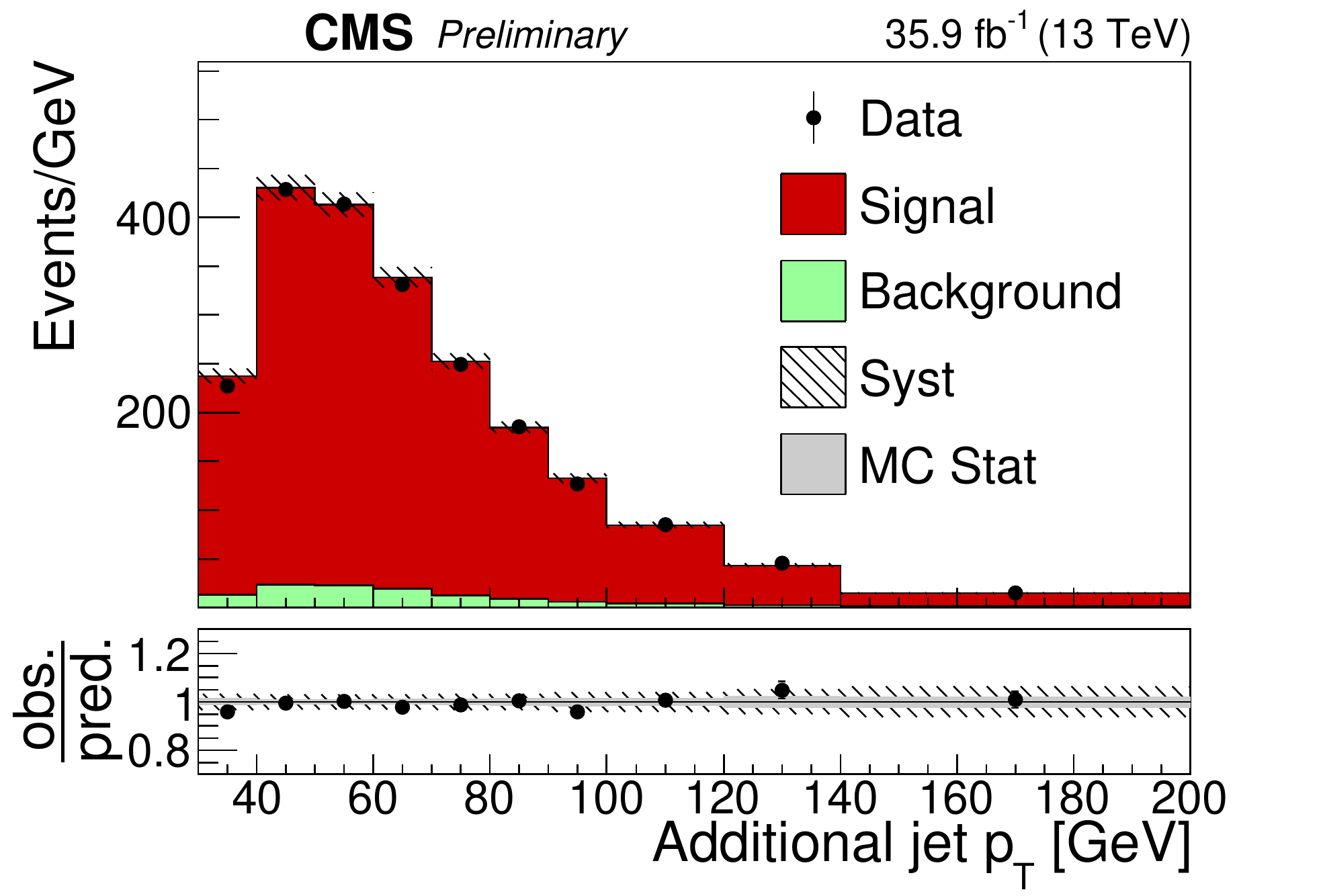}
	\includegraphics[width=0.5\textwidth]{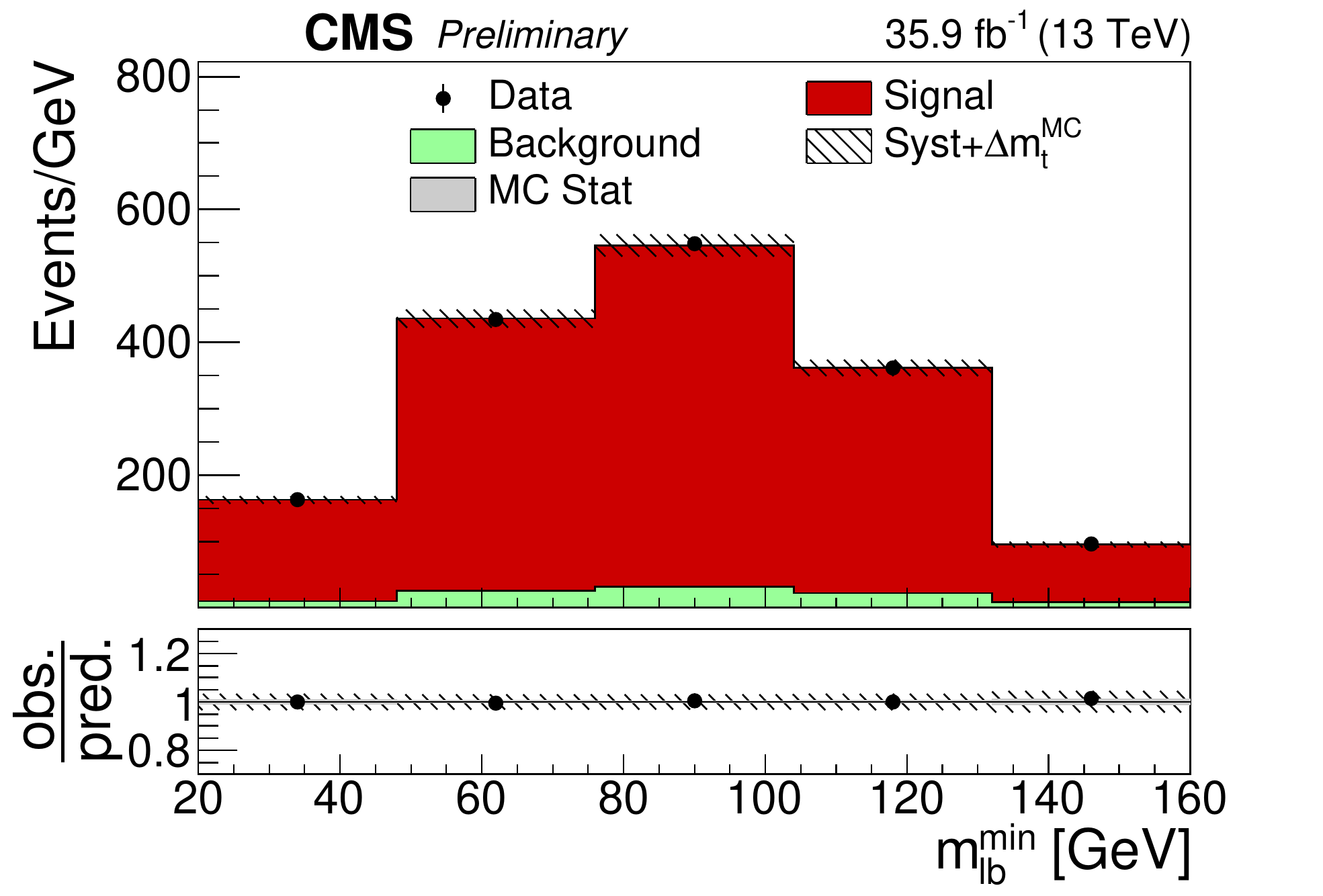}
	\caption{Example of post-fit distributions for the fit of \stt at the fixed \mtmc of 172.5~GeV (left) and for the simultaneous fit of \stt and \mtmc (right). The jet \pt spectrum corresponds to events with one b-tagged jets and two additional (non-b-tagged) jets, whereas the \mlb distribution corresponds to events with one b-tagged jets and one additional jet~\cite{TOP-17-001}.}
	\label{fig:postfit}
\end{figure}

The method consist of a likelihood fit to distributions of final state observables, in bins of jet and b-tagged jet multiplicities. Systematic uncertainties are treated as nuisance parameters and are constrained in the fiducial phase space. Spectra of jet transverse momenta (\pt) are used to constrain the jet energy scale uncertainties, while the \mlb distribution is used to determine \mtmc in the simultaneous fit of \stt and \mtmc (Fig.~\ref{fig:postfit}).  \\

At the fixed value of $\mtmc = 172.5~\GeV$, the visible cross section is measured to be $\sttvis=\resultxsecvismain$. The visible phase space is defined as the events having two leptons of opposite charge with $\pt > 20 \GeV$, at least one of which with $\pt > 25 \GeV$. The extrapolated cross section is found to be $\stt = \resultxsecmain$. The total uncertainty of 4.0\% is dominated by the uncertainties in the integrated luminosity (2.5\%) and the lepton identification efficiencies (2.0\%). %Uncertainties on proton PDF and the modelling and normalization of the tW background also play a relevant role. 
The measured cross section is in good agreement with the theoretical prediction at NNLO+NNLL accuracy $\stttheo = \xsectheo$~\cite{Czakon:2011xx}. \\

In the simultaneous fit of \stt and \mtmc, a total cross section of $\stt = \resultxsectopmass$ is obtained, in good agreement with the result at the fixed \mtmc. The top quark mass in the simulation is measured to be $\mtmc = \resulttopmassMC$, in good agreement with previous measurements by ATLAS and CMS~\cite{sum-plots}. The uncertainty in \mtmc is dominated by jet energy scale uncertainties (0.57~GeV) and the statistical uncertainty of the simulation (0.36~GeV). The latter is estimated by using pseudo-templates, where each bin is varied within its statistical uncertainty. The correlation between different templates is properly taken into account. The fit to the data is repeated several thousand times, and the spread in the best-fit values of \mtmc and \stt is taken as an additional uncertainty. The measured \stt is then used to extract \as and \mt using different PDF sets. Updated results can be found in Ref.~\cite{Sirunyan:2018goh}.

\section{Conclusions}

The most recent measurements of the inclusive \ttbar production cross section by the ATLAS and CMS experiments have been summarized. These include measurements at centre-of-mass energies of 5.02, 8, and 13~TeV in different final states. The first observation of \ttbar production in p-Pb collisions at a nucleon-nucleon centre-of-mass energy of 8.16~TeV has also been illustrated. A new result at $\sqrt{s} = 13~\mathrm{TeV}$ by the CMS Collaboration, obtained using \lumiv of pp collisions data, is presented for the first time at this conference. In this analysis, the \ttbar cross section is determined both for a fixed top quark mass in the simulation of $\mtmc = 172.5~\GeV$, and simultaneously with \mtmc. The cross section is measured with a relative uncertainty of 4.0\%, and good precision in the top quark mass is also achieved.

\end{document}

%% file: econfmacros.tex
\def\beq{\begin{equation}}
\def\eeq#1{\label{#1}\end{equation}}
\def\eeqn{\end{equation}}

\def\beqa{\begin{eqnarray}}
\def\eeqa#1{\label{#1}\end{eqnarray}}
\def\eeqan{\end{eqnarray}}

\let\bar=\overbar

\def\Dslash{\not{\hbox{\kern-4pt $D$}}}
\def\dslash{\not{\hbox{\kern-2pt $\del$}}}

\def\msb{{\bar{\ssstyle M \kern -1pt S}}}

\newcommand{\mumu}{$\mu^+\mu^-$\xspace}
\newcommand{\ee}{$\mathrm{e^+e^-}$\xspace}
\newcommand{\emu}{$\mathrm{e^{\pm}}\mu^{\mp}$\xspace}

\newcommand{\pb}{\mbox{\ensuremath{\,\text{pb}}}\xspace}
\newcommand{\GeV}{\mbox{\ensuremath{\,\text{GeV}}}\xspace}

\newcommand{\stat}{\mbox{\ensuremath{\,\text{(stat)}}}\xspace}
\newcommand{\syst}{\mbox{\ensuremath{\,\text{(syst)}}}\xspace}
\newcommand{\lum}{\mbox{\ensuremath{\,\text{(lum)}}}\xspace}
\newcommand{\ttbar}{\ensuremath{\rm t\bar{t}}\xspace}
\newcommand{\resultxsecmain}{\ensuremath{803  \pm  2 \stat \pm 25 \syst \pm 20 \lum \pb}\xspace}

\newcommand{\resultxsecvismain}{\ensuremath{25.61 \pm 0.05 \stat \pm 0.75 \syst \pm 0.64 \lum \pb }\xspace}

\newcommand{\xsectheo}{\ensuremath{832 \pm^{20}_{29} {\rm (scale)} \pm 35 ~({\rm PDF}+\as) \pb}\xspace}

\newcommand{\resultxsectopmass}{\ensuremath{815 \pm  2 \stat \pm 29 \syst \pm 20 \lum \pb}\xspace}
\newcommand{\resulttopmassMC}{\ensuremath{172.33 \pm  0.14 \stat \pm^{0.66}_{0.72} \syst \GeV}\xspace}

\newcommand{\lumiv}{35.9 fb$^{-1}$\xspace}

\newcommand{\as}{\ensuremath{\alpha_\mathrm{S}}\xspace}

\newcommand{\stt}{\ensuremath{\sigma_\mathrm{t\bar{t}}}\xspace}
\newcommand{\sttvis}{\ensuremath{\sigma_{\ttbar}^{\mathrm{vis}}}\xspace}
\newcommand{\pt}{\ensuremath{\mathrm{p_T}}\xspace}
\newcommand{\stttheo}{\ensuremath{\sigma_{\ttbar}^{\mathrm{theo}}\xspace}}

\newcommand{\mtmc}{\ensuremath{m_\mathrm{t}^{\mathrm{MC}}}\xspace}

\newcommand{\mt}{\ensuremath{m_\mathrm{t}}\xspace}

\newcommand{\mlb}{\ensuremath{m_\mathrm{\ell b}^{\mathrm{min}}}\xspace}

%% file: TOP2018_ATLAS_CMS_xsec.bbl
\begin{thebibliography}{99}

%%
%%  bibliographic items can be constructed using the LaTeX format in SPIRES:
%%    see    http://www.slac.stanford.edu/spires/hep/latex.html
%%  SPIRES will also supply the CITATION line information; please include it.
%%

%\cite{Aad:2008zzm}
\bibitem{Aad:2008zzm}
ATLAS Collaboration,
%``The ATLAS Experiment at the CERN Large Hadron Collider,''
JINST {\bf 3} (2008) S08003.
%doi:10.1088/1748-0221/3/08/S08003
%%CITATION = doi:10.1088/1748-0221/3/08/S08003;%%
%6927 citations counted in INSPIRE as of 17 Dec 2018

%\cite{Chatrchyan:2008aa}
\bibitem{Chatrchyan:2008aa}
CMS Collaboration,
%``The CMS Experiment at the CERN LHC,''
JINST {\bf 3} (2008) S08004.
%doi:10.1088/1748-0221/3/08/S08004
%%CITATION = doi:10.1088/1748-0221/3/08/S08004;%%
%5716 citations counted in INSPIRE as of 17 Dec 2018

\bibitem{Aaboud:2017cgs}
ATLAS Collaboration,
%``Measurement of the inclusive and fiducial $t\bar{t}$ production cross-sections in the lepton+jets channel in $pp$ collisions at $\sqrt{s} = 8$ TeV with the ATLAS detector,''
Eur.\ Phys.\ J.\ C {\bf 78} (2018) 487
%doi:10.1140/epjc/s10052-018-5904-z
%[arXiv:1712.06857 [hep-ex]].
%%CITATION = doi:10.1140/epjc/s10052-018-5904-z;%%
%5 citations counted in INSPIRE as of 23 Nov 2018

%\cite{Sirunyan:2017uhy}
\bibitem{Sirunyan:2017uhy}
CMS Collaboration,
%``Measurement of the $t \bar t$ production cross section using events with one lepton and at least one jet in pp collisions at $\sqrt{s}$  = 13 TeV,''
JHEP {\bf 1709} (2017) 051
%doi:10.1007/JHEP09(2017)051
%[arXiv:1701.06228 [hep-ex]].
%%CITATION = doi:10.1007/JHEP09(2017)051;%%
%38 citations counted in INSPIRE as of 23 Nov 2018

%\cite{Aaboud:2016pbd}
\bibitem{Aaboud:2016pbd}
ATLAS Collaboration,
%``Measurement of the $t\bar{t}$ production cross-section using $e\mu$ events with b-tagged jets in pp collisions at $\sqrt{s}$=13 TeV with the ATLAS detector,''
Phys.\ Lett.\ B {\bf 761} (2016) 136
%Erratum: [Phys.\ Lett.\ B {\bf 772} (2017) 879]
%doi:10.1016/j.physletb.2016.08.019, 10.1016/j.physletb.2017.09.027
%[arXiv:1606.02699 [hep-ex]].
%%CITATION = doi:10.1016/j.physletb.2016.08.019, 10.1016/j.physletb.2017.09.027;%%
%91 citations counted in INSPIRE as of 23 Nov 2018

%\cite{Aaboud:2016zpd}
\bibitem{Aaboud:2016zpd}
ATLAS Collaboration,
%``Measurements of top-quark pair to $Z$-boson cross-section ratios at $\sqrt s = 13, 8, 7$ TeV with the ATLAS detector,''
JHEP {\bf 1702} (2017) 117
%doi:10.1007/JHEP02(2017)117
%[arXiv:1612.03636 [hep-ex]].
%%CITATION = doi:10.1007/JHEP02(2017)117;%%
%20 citations counted in INSPIRE as of 23 Nov 2018

%\cite{Sirunyan:2017ule}
\bibitem{Sirunyan:2017ule}
CMS Collaboration,
%``Measurement of the inclusive $ \mathrm{t}\overline{\mathrm{t}} $ cross section in pp collisions at $ \sqrt{s}=5.02 $ TeV using final states with at least one charged lepton,''
JHEP {\bf 1803} (2018) 115
%doi:10.1007/JHEP03(2018)115
%[arXiv:1711.03143 [hep-ex]].
%%CITATION = doi:10.1007/JHEP03(2018)115;%%
%11 citations counted in INSPIRE as of 23 Nov 2018

%\cite{Sirunyan:2017xku}
\bibitem{Sirunyan:2017xku}
CMS Collaboration,
%``Observation of top quark production in proton-nucleus collisions,''
Phys.\ Rev.\ Lett.\  {\bf 119} (2017) no.24,  242001
%doi:10.1103/PhysRevLett.119.242001
%[arXiv:1709.07411 [nucl-ex]].
%%CITATION = doi:10.1103/PhysRevLett.119.242001;%%
%13 citations counted in INSPIRE as of 23 Nov 2018

\bibitem{TOP-17-001}
CMS Collaboration, CMS-PAS-TOP-17-001 (2018) \\ https://cds.cern.ch/record/2639182

%\cite{Sirunyan:2018goh}
\bibitem{Sirunyan:2018goh}
CMS Collaboration,
%``Measurement of the $\mathrm{t}\overline{\mathrm{t}}$ production cross section, the top quark mass, and the strong coupling constant using dilepton events in pp collisions at $\sqrt{s}=$ 13 TeV,''
arXiv:1812.10505.
%%CITATION = ARXIV:1812.10505;%%

\bibitem{Czakon:2011xx}
M.~Czakon and A.~Mitov,
%``Top++: A Program for the Calculation of the Top-Pair Cross-Section at Hadron Colliders,''
Comput.\ Phys.\ Commun.\  {\bf 185} (2014) 2930
%doi:10.1016/j.cpc.2014.06.021
%[arXiv:1112.5675 [hep-ph]].
%%CITATION = doi:10.1016/j.cpc.2014.06.021;%%
%946 citations counted in INSPIRE as of 11 Dec 2018

\bibitem{sum-plots}
twiki.cern.ch/twiki/bin/view/LHCPhysics/LHCTopWGSummaryPlots

\end{thebibliography}
